\documentclass[onecolumn,english,aps,prL,amsmath,amssymb,superscriptaddress]{revtex4}
\usepackage[latin9]{inputenc}
\usepackage{color}
\usepackage{amstext}
\usepackage{amssymb}
\usepackage{graphicx}
\usepackage[multiple]{footmisc}

\makeatletter
\@ifundefined{textcolor}{}
{%
 \definecolor{BLACK}{gray}{0}
 \definecolor{WHITE}{gray}{1}
 \definecolor{RED}{rgb}{1,0,0}
 \definecolor{GREEN}{rgb}{0,1,0}
 \definecolor{BLUE}{rgb}{0,0,1}
 \definecolor{CYAN}{cmyk}{1,0,0,0}
 \definecolor{MAGENTA}{cmyk}{0,1,0,0}
 \definecolor{YELLOW}{cmyk}{0,0,1,0}
}

\usepackage{color}

\newcommand{\beq}{\begin{eqnarray}}
\newcommand{\eeq}{\end{eqnarray}}

\makeatother
\usepackage{babel}

\begin{document}

\title{Consecutive topological phase transitions and colossal magnetoresistance in a magnetic topological semimetal}


\author{Feng Du\footnote{These authors contributed equally to this work.}}
\affiliation{Center for Correlated Matter and Department of Physics, Zhejiang University, Hangzhou 310058, China}

\author{Lin Yang$\rm ^*$}
\email{lyang@hdu.edu.cn}
\affiliation{College of Materials and Environmental Engineering, Hangzhou Dianzi University, Hangzhou 310018, China}

\author{Zhiyong Nie}
\affiliation{Center for Correlated Matter and Department of Physics, Zhejiang University, Hangzhou 310058, China}

\author{Ninghua Wu}
\affiliation{Center for Correlated Matter and Department of Physics, Zhejiang University, Hangzhou 310058, China}

\author{Yong Li}
\affiliation{Beijing National Laboratory for Condensed Matter Physics and Institute of Physics, Chinese Academy of Sciences, Beijing 100190, China}

\author{Shuaishuai Luo}
\affiliation{Center for Correlated Matter and Department of Physics, Zhejiang University, Hangzhou 310058, China}

\author{Ye Chen}
\affiliation{Center for Correlated Matter and Department of Physics, Zhejiang University, Hangzhou 310058, China}
\affiliation  {Zhejiang Province Key Laboratory of Quantum Technology and Device, Department of Physics, Zhejiang University, Hangzhou 310058, China}

\author{Dajun Su}
\affiliation{Center for Correlated Matter and Department of Physics, Zhejiang University, Hangzhou 310058, China}

\author{Michael Smidman}
\affiliation{Center for Correlated Matter and Department of Physics, Zhejiang University, Hangzhou 310058, China}
\affiliation  {Zhejiang Province Key Laboratory of Quantum Technology and Device, Department of Physics, Zhejiang University, Hangzhou 310058, China}

\author{Youguo Shi}
\affiliation{Beijing National Laboratory for Condensed Matter Physics and Institute of Physics, Chinese Academy of Sciences, Beijing 100190, China}

\author{Chao Cao}

\affiliation{Center for Correlated Matter and Department of Physics, Zhejiang University, Hangzhou 310058, China}

\author{Frank Steglich}
\affiliation{Center for Correlated Matter and Department of Physics, Zhejiang University, Hangzhou 310058, China}
\affiliation{Max Planck Institute for Chemical Physics of Solids, 01187 Dresden, Germany}

\author{Yu Song}
\email{yusong$_$phys@zju.edu.cn}
\affiliation{Center for Correlated Matter and Department of Physics, Zhejiang University, Hangzhou 310058, China}
\affiliation  {Zhejiang Province Key Laboratory of Quantum Technology and Device, Department of Physics, Zhejiang University, Hangzhou 310058, China}

\author{Huiqiu Yuan}
\email{hqyuan@zju.edu.cn}
\selectlanguage{english}%
\affiliation{Center for Correlated Matter and Department of Physics, Zhejiang University, Hangzhou 310058, China}
\affiliation  {Zhejiang Province Key Laboratory of Quantum Technology and Device, Department of Physics, Zhejiang University, Hangzhou 310058, China}
\affiliation  {State Key Laboratory of Silicon Materials, Zhejiang University, Hangzhou 310058, China}

\begin{abstract}
The combination of magnetic symmetries and electronic band topology provides a promising route for realizing topologically nontrivial quasiparticles, and the manipulation of magnetic structures may enable the switching between topological phases, with the potential for achieving functional physical properties. 
Here, we report measurements of the electrical resistivity of EuCd$_2$As$_2$ under pressure, which show an intriguing insulating dome at pressures between $p_{\rm c1}\sim1.0$~GPa and $p_{\rm c2}\sim2.0$~GPa, situated between two regimes with metallic transport.
The insulating state can be fully suppressed by a small magnetic field, leading to a colossal negative magnetoresistance on the order of $10^5$\%, accessible via a modest field of $\sim0.2$~T. First-principles calculations reveal that the dramatic evolution of the resistivity under pressure is due to consecutive transitions of EuCd$_2$As$_2$ from a magnetic topological insulator to a trivial insulator, and then to a Weyl semimetal, with the latter resulting from a pressure-induced change in the magnetic ground state. Similarly, the colossal magnetoresistance results from a field-induced polarization of the magnetic moments, transforming EuCd$_2$As$_2$ from a trivial insulator to a Weyl semimetal. These findings underscore weak magnetic exchange couplings and spin anisotropy as ingredients for discovering tunable magnetic topological materials with desirable functionalities. 

\end{abstract}

\maketitle

\section*{Introduction}

Whereas the electronic band topology in nonmagnetic topological materials is determined by the crystal symmetry and the relative strength of spin-orbit coupling, the magnetic ground state also plays a decisive role in magnetic topological materials \cite{Tokura2019,zou2019study,Watanabe2018,Xu2020,Elcoro2021,Wang2021magnetic}. The concurrence of magnetism and electronic band topology leads to novel states of matter, including Weyl fermions in centrosymmetric systems \cite{Kuroda2017,Liu2018}, magnetic topological insulators \cite{Otrokov2019}, quantum anomalous Hall effect \cite{deng2020quantum}, and axion insulators \cite{Sekine2021}. The dependence of band topology on the magnetic ground state grants access to topological phase transitions by manipulation of the latter, and opens up a route for discovering functional properties in magnetic topological materials.

$f$-electron materials provide a fertile setting for discovering emergent physics at the intersection of magnetism, electron correlations and topology, including magnetic topological insulators \cite{Hua2018,Wang2019,Xu2019,Li2019}, as well as topological spin textures \cite{Kurumaji2019,Hirschberger2019,Kaneko2019,Puphal2020} and strongly correlated topological phases such as topological Kondo insulators and semimetals \cite{Dzero2010,Kim2013,xu2014,Guo2018}. 
Among the $f$-electron magnetic topological materials, EuCd$_2$As$_2$ exhibits a single Dirac cone at the Fermi level well inside the paramagnetic state, offering an ideal platform for realizing exotic quasiparticles \cite{Hua2018,Niu2019,Xu2021}. Below $T_{\rm N}\approx9$~K, the Eu$^{2+}$ moments are oriented in the $ab$-plane and form A-type antiferromagnetic (AFM) order, with antiferromagnetically stacked ferromagnetic (FM) Eu$^{2+}$ layers \cite{Rahn2018}. With such a magnetic ground state, a gap opens at the Dirac point and the system may be a small-gap magnetic topological insulator (MTI) \cite{Wang2019,Ma2020}. A fully polarized state along the $c$-axis can be readily obtained upon applying a magnetic field, giving rise to an ideal single pair of Weyl points \cite{Wang2019,Soh2019}. A Weyl state is also suggested for the paramagnetic phase slightly above $T_{\rm N}$ based on photoemission measurements, resulting from a proliferation of quasi-static in-plane ferromagnetic fluctuations \cite{Ma2019,Soh2020}. A plethora of additional topological states, including magnetic Dirac fermions, axion insulator, and higher-order topological insulator, are also expected given the appropriate magnetic ground state \cite{Ma2020}. Moreover, in addition to becoming fully polarized under modest in-plane or $c$-axis magnetic fields \cite{Rahn2018}, 
a FM ground state can be stabilized by changes in the synthesis protocol \cite{Jo2020} or applying hydrostatic pressure \cite{gati2021pressureinduced}, demonstrating highly-tunable magnetism. These behaviors highlight the proximity of competing magnetic ground states in EuCd$_2$As$_2$, which can be harnessed to manipulate its electronic topology.

Here, by carrying out resistivity measurements on EuCd$_2$As$_2$ single crystals under hydrostatic pressure, we find an insulating dome within a small pressure window between $p_{\rm c1}\sim1.0$~GPa and $p_{\rm c2}\sim2.0$~GPa, straddled by two regimes ($p<p_{\rm c1}$ and $p>p_{\rm c2}$) with metallic transport. The insulating state is easily suppressed by a magnetic field (0.2~T for $H\perp c$), leading to a remarkable colossal negative magnetoresistance (MR) which reaches $\sim10^5$\% at 0.3~K,
and can likely be enhanced upon further cooling. 
Based on first-principles calculations, these experimental observations may arise from two topological phase transitions tuned by magnetism. Namely, a transition from an AFM topological insulator to an AFM trivial insulator (TrI) at $p_{\rm c1}\sim1.0$~GPa, and a transition from an AFM TrI to a FM Weyl semimetal (WSM) at $p_{\rm c2}\sim2.0$~GPa, triggered by a pressure-induced AFM-FM transition of the magnetic ground state. A similar mechanism accounts for the suppression of the TrI state under applied magnetic field, which polarizes the AFM state and transforms the system to a WSM, leading to the colossal MR. 
These findings demonstrate that the combination of Dirac fermions and proximate magnetic ground states provides a route for discovering tunable topological quantum materials with functional properties.

\begin{figure}
	\includegraphics[width=0.7\columnwidth]{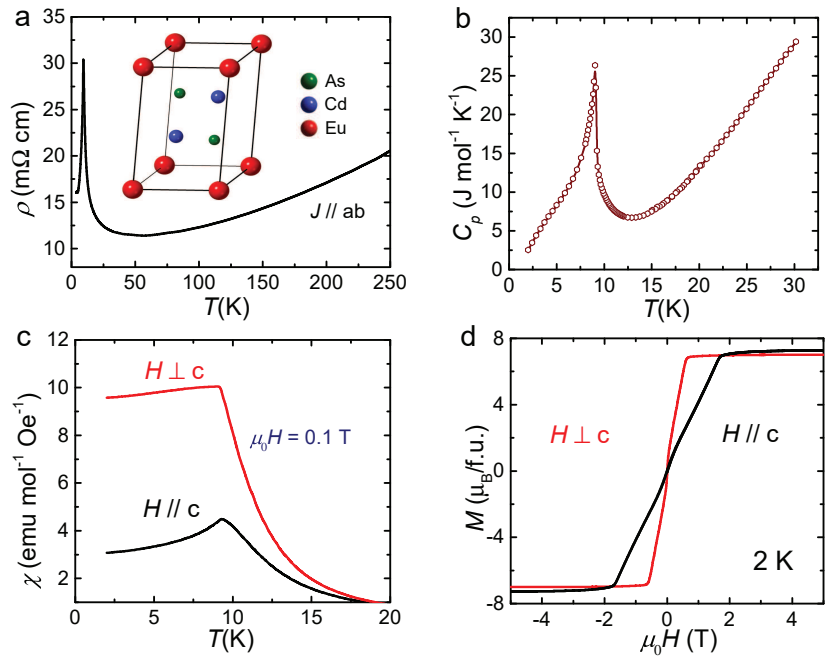} \protect\caption{{\bf Physical properties of EuCd$_2$As$_2$ at ambient pressure.} (a) Temperature dependence of the resistivity $\rho(T)$ in the $ab$-plane. The primitive unit cell is shown in the inset, with Eu$^2+$ ions forming a triangular lattice in the $ab$-plane; (b) Total specific heat $C_{\rm p}(T)$; (c) Magnetic susceptibility $\chi(T)$, measured in an applied field of $\mu_0 H=0.1$~T; (d) The field-dependence of the magnetization $M(H)$, measured for two field directions at 2~K after zero-field cooling.}
	\label{Fig_basic}
\end{figure}

\section*{Results}
\subsection*{Metallic-insulating-metallic evolution of electrical transport under pressure}
EuCd$_2$As$_2$ crystallizes in a centrosymmetric trigonal structure (unit cell volume $V=125.03$~\AA$^{3}$ \cite{sample}), with planes of Eu$^{2+}$ that form triangular lattices [inset of Fig.~\ref{Fig_basic}(a)] separated by layers of Cd-As tetrahedra networks \cite{Artmann1996}. Whereas the FM alignment of magnetic moments within the $ab$-plane is robust, the stacking of these FM planes can be tuned between AFM or FM order along the $c$-axis, resulting in an overall A-type AFM or FM ground state \cite{Rahn2018,Krishna2018,Jo2020}. 
Given the proximity between these magnetic ground states, the physical properties of our EuCd$_2$As$_2$ samples were carefully characterized at ambient pressure, with the results summarized in Fig.~\ref{Fig_basic}. Clear peaks in both the resistivity $\rho(T)$ and specific heat $C_{\rm p}(T)$ are observed around $T_{\rm N}\approx9$~K [Figs.~\ref{Fig_basic}(a) and (b)].
Measurements of the magnetic susceptibility under a small field of 0.1~T [Fig.~\ref{Fig_basic}(c)] suggest an AFM ground state with an easy $ab$-plane, in agreement with resonant x-ray scattering measurements \cite{Rahn2018}. Magnetization measurements reveal saturation fields (saturated moments) of about $1.7$~T (7.28~$\mu_{\rm B}$/Eu) for $H\parallel c$ and $0.7$~T (7.01~$\mu_{\rm B}$/Eu) for $H\perp c$ [Fig.~\ref{Fig_basic}(d)]. The saturated moments are close to the ideal value of 7.0~$\mu_{\rm B}$ for Eu$^{2+}$ ions, indicating localized magnetism. These characterizations establish that the EuCd$_2$As$_2$ samples used in this study exhibit an A-type AFM ground state below $T_{\rm N}\approx9$~K at ambient pressure, similar to previous reports \cite{Wang2016,Rahn2018,gati2021pressureinduced}.

\begin{figure}
	\includegraphics[width=0.7\columnwidth]{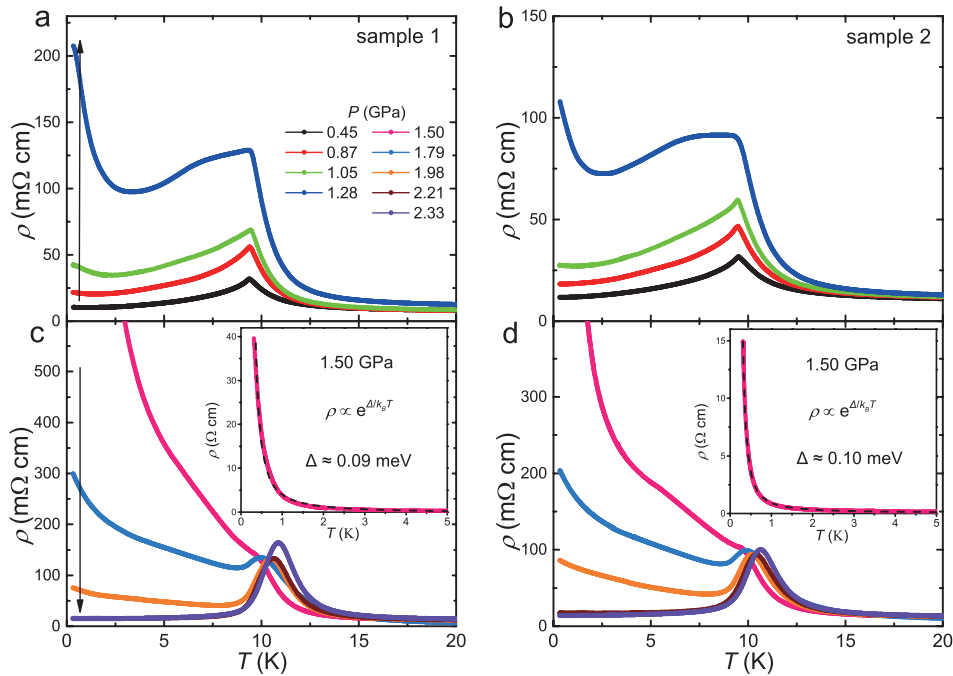} \protect\caption{{\bf Metallic-insulating-metallic evolution of electrical transport under applied pressure.} Temperature dependence of the $ab$-plane resistivity $\rho(T)$ of EuCd$_2$As$_2$ under various pressures from 0.45~GPa to 1.28~GPa, for samples (a) \#1 and (b) \#2. The corresponding $\rho(T)$ under pressures from 1.50~GPa to 2.21~GPa are shown in panels (c) and (d). The insets of (c) and (d) show $\rho(T)$ under 1.50~GPa and the corresponding fits to $\rho(T)-\rho_0\propto e^{\frac{\Delta}{k_{\rm B} T}}$, where $\rho_0$ is a temperature-independent contribution to the resistivity. The vertical arrows in (a) and (c) mark the direction of increasing pressure.}
	\label{Fig_rho}
\end{figure}


In order to track the pressure-evolution of the ground state properties of EuCd$_2$As$_2$, electrical resistivity measurements were carried out on four samples under pressures up to 2.50~GPa and temperatures down to 0.3~K. The results for two samples \#1 and \#2 are presented in Fig.~\ref{Fig_rho}, and the others are shown in the Supplementary Materials (see Fig.~S1). All samples exhibit consistent behaviors under pressure.

For pressures up to 0.87~GPa [Figs.~\ref{Fig_rho}(a) and (b)], $\rho(T)$ is qualitatively similar to that at ambient pressure, with a sharp peak around $T_{\rm N}$ and metallic behavior at low temperatures [$\rho(0.3~{\rm K})\lesssim20$~m$\Omega\cdot$cm]. Upon increasing the pressure to 1.05~GPa, an additional subtle upturn appears when cooling below $\sim2$~K. The upturn in $\rho(T)$ at low temperatures becomes more prominent under 1.28~GPa, and is maximized around 1.50~GPa [Figs.~\ref{Fig_rho}(c) and (d)], with $\rho(0.3~{\rm K})$ reaching $\approx40$~$\Omega\cdot$cm and $\approx15$~$\Omega\cdot$cm, for samples \#1 and \#2 respectively [insets in Figs.~\ref{Fig_rho}(c) and (d)]. The temperature-dependence of the resistivity under 1.50~GPa can be captured by $\rho(T)-\rho_0\propto\exp(\frac{\Delta}{k_{\rm B}T})$, where $\rho_0$ is a temperature-independent contribution and $\Delta$ represents an energy gap. This indicates that the upturn is associated with an electronic gap, suggesting a metal-insulator transition in EuCd$_2$As$_2$ across $p_{\rm c1}\sim1.0$~GPa. By fitting $\rho(T)$ of all measured samples for $T<5$~K, $\Delta$ is consistently found to be $\approx0.07-0.10$~meV at 1.50~GPa (see Fig.~S1 and Section~S1 in the Supplementary Materials), suggesting that $\Delta$ does not have a significant sample dependence.

\begin{figure}
	\includegraphics[width=0.7\columnwidth]{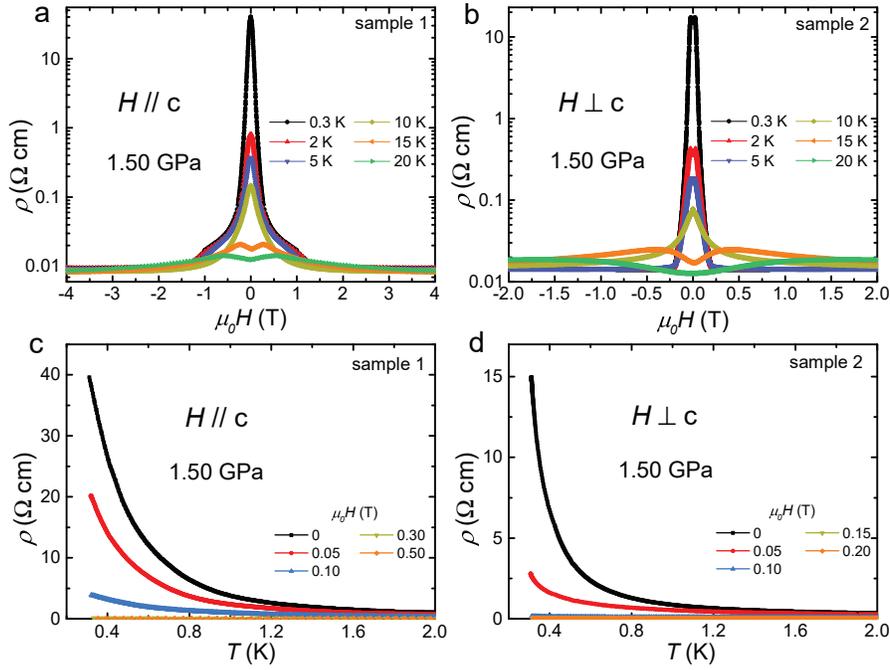} \protect\caption{{\bf Colossal magnetoresistance in EuCd$_2$As$_2$ under 1.50~GPa.} Field-dependence of the resistivity $\rho(H)$ at various temperatures, for fields (a) along the $c$-axis, and (b) in the $ab$-plane, plotted on a log scale. Temperature-dependence of the resistivity $\rho(T)$ under various magnetic fields along (c) the $c$-axis, and in (d) the $ab$-plane.}
	\label{Fig_MR}
\end{figure}

Upon further increasing the pressure, the upturn in $\rho(T)$ weakens under 1.79~GPa and disappears for $p\gtrsim2.21$~GPa, at which $\rho(0.3~{\rm K})$ reverts back to $\lesssim20$~m$\Omega\cdot$cm. Such an evolution points to the restoration of metallic transport above $p_{\rm c2}\sim2.0$~GPa, and experimentally establishes the presence of an insulating dome within the magnetically ordered state for $p_{\rm c1}<p<p_{\rm c2}$.
Since A-type AFM order is present in EuCd$_2$As$_2$ up to $\sim p_{\rm c2}$, as indicated by the response of the resistivity to small magnetic fields (see Fig.~S2 and Section~S2 in the Supplementary Materials) and previous transport and $\mu$SR measurements \cite{gati2021pressureinduced}, the insulating state appears tied to the AFM order.

\section*{Colossal magnetoresistance in the insulating state}

To further elucidate the connection between the magnetic order and the insulating state under pressure, magnetoresistance measurements were carried out under 1.50~GPa for samples~\#1 and \#2, and the results are shown in Fig.~\ref{Fig_MR}. Since the saturation field of EuCd$_2$As$_2$ is small [Fig.~\ref{Fig_basic}(d)], and further reduces with increasing pressure \cite{gati2021pressureinduced}, it is straightforward to assume that the A-type AFM order of EuCd$_2$As$_2$ can be polarized to achieve a FM state using accessible magnetic fields. If the insulating state indeed relies on the presence of A-type AFM order, it should be suppressed when the AFM state becomes fully polarized.
 
At 1.5~GPa, with increasing field both along the $c$-axis and in the $ab$-plane, $\rho(0.3~{\rm K})$ reduces by around three orders of magnitude, and settles to values less than $20$~m$\Omega\cdot$cm, as shown in Figs.~\ref{Fig_MR}(a) and (b). This transformation from insulating to metallic electrical transport is similar to the pressure-induced evolution across $p_{\rm c2}$. Quantitatively, the MR [defined as $[\rho(H)-\rho(0)]/\rho(0)$] at 0.3~K reaches $-3.78\times10^5$\% for sample~\#1 with $H\parallel c$ [Fig.~\ref{Fig_MR}(a)], and $-1.05\times10^5$\% for sample~\#2 with $H\perp c$ [Fig.~\ref{Fig_MR}(b)]. The values of the MR in EuCd$_2$As$_2$ are large even when compared to the colossal MR in the manganites \cite{vonHelmolt1993,Jin1994,Salamon2001}, and are even more striking considering that they are achieved with relatively small fields [Table~\ref{table_MR}]. The fields at which the MR saturates are $\mu_0H\sim1.5$~T for $H\parallel c$ and $\mu_0H\sim0.2$~T for $H\perp c$, which are in excellent agreement with the saturation fields \cite{gati2021pressureinduced}, and indicates that the MR is associated with polarization of the Eu$^{2+}$ moments. At higher temperatures, the resistivity in zero field is substantially reduced, whereas the resistivity under $\left|\mu_0H\right|\gtrsim2$~T remains similar to that at 0.3~K. This leads to a weakening of the MR upon warming, originating from a decrease in the zero-field resistivity with increasing temperature.


The colossal negative MR can also be seen from the temperature dependence of the resistivity under various fields for samples~\#1 and \#2 at 1.50~GPa, as shown in Figs.~\ref{Fig_MR}(c) and (d). Similar to the results in Figs.~\ref{Fig_rho}(c) and (d), where the insulating state is suppressed by increasing pressure above $p_{\rm c2}$, modest magnetic fields also efficiently suppress the insulating state. In both cases, the A-type AFM phase is destabilized and superseded by a FM state, respectively due to an AFM-FM transition at $p_{\rm c2}\sim2.0$~GPa, or the full polarization of Eu$^{2+}$ moments by an applied field.  
While a sizable MR is also seen in EuCd$_2$As$_2$ at ambient pressure near $T_{\rm N}$, it should be emphasized that the colossal MR we have uncovered in the insulating state of EuCd$_2$As$_2$ is distinct. Whereas the former is associated with critical fluctuations near a magnetic transition with similar behaviors detected in a number of Eu-based compounds \cite{Oliver1972,Shapira1972,Wang2021}, the MR shown in Fig.~\ref{Fig_MR} is linked to a field-induced transition from an insulating state with a small electronic gap $\Delta$ to a metallic state, and persists down to at least 0.3~K. Therefore, although the MR in our measurements may contain contributions related to critical fluctuations near the magnetic transition temperature, the MR at low temperatures (e.g. $T<2$~K) clearly has a different, and thus far unreported mechanism.  

\begin{table}[tb]
	\caption{Magnetic fields needed to achieve various values of magnetoresistance in EuCd$_2$As$_2$ under 1.50~GPa at 0.3~K}
	\label{table_MR}
	\begin{tabular}{|c |c |c |c |c|}
		\hline	Field direction & MR ($-10^5$\%) & MR ($-10^4$\%) & MR ($-10^3$\%) & MR ($-10^2$\%)\\
		\hline	H $\parallel$ c (sample~\#1) & 0.42~T& 0.18~T & 0.10~T & 0.05~T \\
		\hline	H $\perp$ c (sample~\#2) & 0.22~T& 0.10~T & 0.06~T & 0.04~T \\
		\hline
	\end{tabular}
\end{table}

From Figs.~\ref{Fig_MR}(c) and (d), it can be seen that the low-temperature MR gradually decreases with increasing temperature. On the flip side, this suggests that the MR should continue to increase as the temperature is lowered below 0.3~K. In fact, since the resistivity tends to infinity for an insulator when the temperature approaches absolute zero, the MR could become substantially larger (possibly by orders of magnitude) than what we observed at 0.3~K. This consideration makes the colossal negative MR in EuCd$_2$As$_2$ of potential interest for ultra-low-temperature applications, especially ones at millikelvin temperatures.

 


\subsection*{Evolution of the electronic structure under pressure}

\begin{figure}
	\includegraphics[width=0.9\columnwidth]{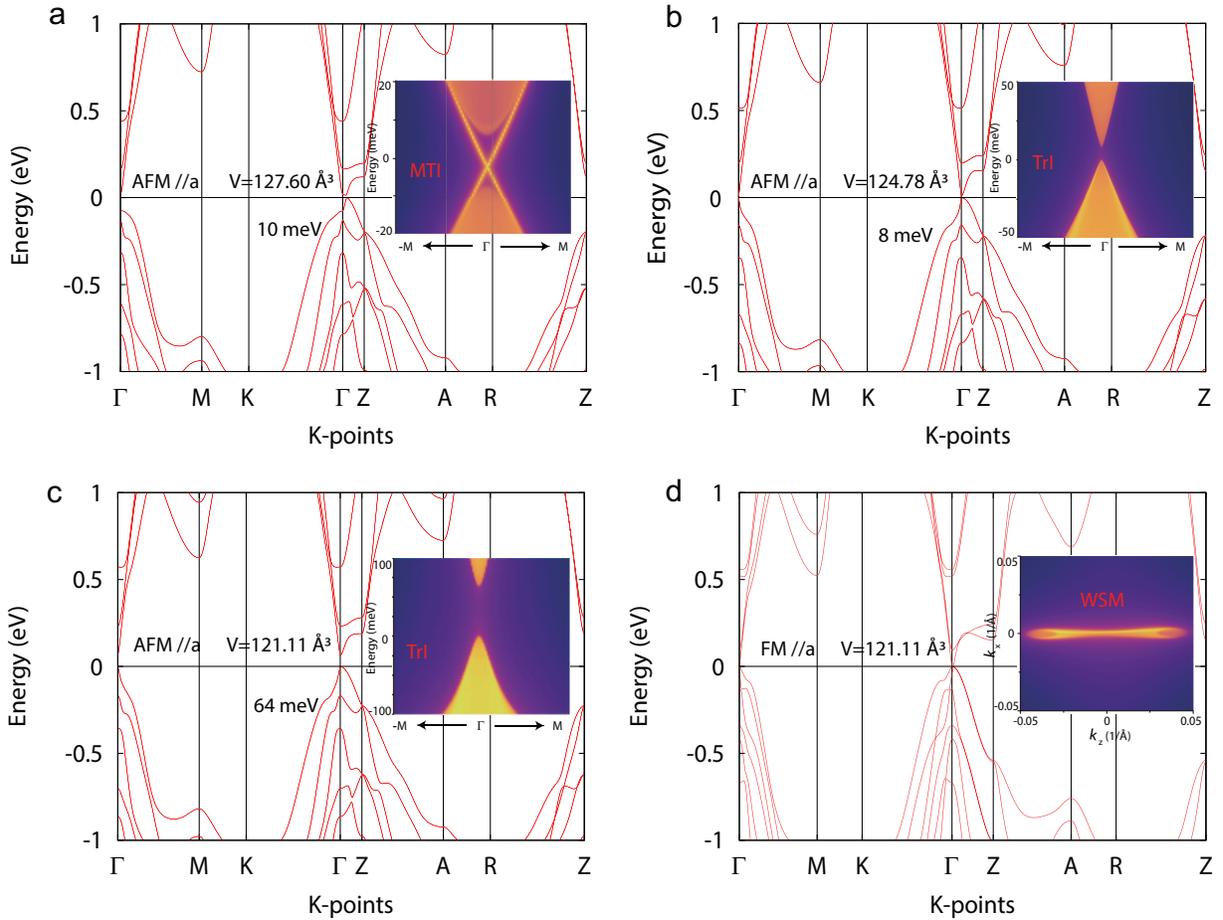} \protect\caption{{\bf Calculated band structures of EuCd$_2$As$_2$ tuned by the magnetic ground state and pressure.} The electronic structure of A-type AFM EuCd$_2$As$_2$ with moments in the $ab$-plane, for unit cell volumes of (a) 127.60~{\AA}$^3$, (b) 124.78~{\AA}$^3$, and (c) 121.11~{\AA}$^3$. (d) The electronic structure for FM EuCd$_2$As$_2$ with moments in the $ab$-plane, with a unit cell volume of 121.11~{\AA}$^3$. Depending on the magnetic ground state and pressure (unit cell volume), EuCd$_2$As$_2$ may be a magnetic topological insulator (MTI), a trivial insulator (TrI), or a Weyl semimetal (WSM). The values of the energy gaps for the MTI and TrI states are indicated in (a)-(c). The insets in (a)-(c) show calculated surface states along $-M\Gamma M$, with clear topologically-protected surface states in (a). The inset in (d) shows the surface Fermi arc along $k_z$ in the $k_x-k_z$-plane.}
	\label{Fig_DFT}
\end{figure}

The experimentally observed insulating dome is difficult to understand without considering magnetic order, and the colossal MR points to a strong coupling between electrical transport, which is dictated by the electronic topology, and magnetism.
To gain insight into the origin of these experimental observations, the electronic structures of EuCd$_2$As$_2$ were calculated for different magnetic ground states and at various pressures, using density functional theory (DFT), with results shown in Fig.~\ref{Fig_DFT}. Applied pressures in DFT calculations are gauged via the unit cell volume $V$, with increasing pressure corresponding to a decrease in $V$. For the A-type AFM state with moments in the $ab$-plane found experimentally \cite{Rahn2018}, EuCd$_2$As$_2$ is a MTI with a 10~meV bulk gap and topologically-protected surface states when $V=127.60$~{\AA}$^3$ [Fig.~\ref{Fig_DFT}(a)]. While the bulk of a MTI is insulating, topologically-protected surface states that cross the Fermi level [inset in Fig.~\ref{Fig_DFT}(a)] give rise to metallic conduction, which accounts for the metallic conduction seen experimentally for $p<p_{\rm c1}$.

Within the A-type AFM state of EuCd$_2$As$_2$, upon increasing pressure such that $V=124.78$~{\AA}$^3$, EuCd$_2$As$_2$ evolves to become a TrI with an 8~meV trivial band gap [Fig.~\ref{Fig_DFT}(b)], where relative to the MTI state, topologically-protected surface states disappear due to an increase in the electronic hopping relative to spin-orbit coupling, which suppresses band inversion. When $V$ becomes further compressed from 124.78~{\AA}$^3$ to 121.11~{\AA}$^3$, the trivial band gap increases from $8$~meV to 64~meV [Fig.~\ref{Fig_DFT}(c)]. For a trivial insulator, its surface states are also gapped [insets in Fig.~\ref{Fig_DFT}(b) and (c)], similar to the bulk, so that the system as a whole lacks mobile carriers. These calculations show that within the A-type AFM state of EuCd$_2$As$_2$, the transition from a MTI to a TrI upon increasing pressure provides an explanation for the experimentally observed evolution of electrical transport across $p_{\rm c1}$.  

While AFM EuCd$_2$As$_2$ with $V=121.11$~{\AA}$^3$ is a TrI [Fig.~\ref{Fig_DFT}(c)], for a FM ground state with in-plane moments under the same pressure (same $V$), EuCd$_2$As$_2$ is a WSM with bulk Weyl points and a surface Fermi arc [Fig.~\ref{Fig_DFT}(d) and its inset]. Such a FM state occurs experimentally when the moments are fully polarized by an $ab$-plane field or when $p>p_{\rm c2}$, with both the Weyl points and the Fermi arc contributing to metallic transport. A FM state with moments along the $c$-axis is also a WSM for $V=121.11$~{\AA}$^3$ (see Section~S3 and Figs.~S3 in the Supplementary Materials), which is realized in the fully polarized state when $H\parallel c$. The DFT results in Figs.~\ref{Fig_DFT}(c) and (d) indicate that for AFM EuCd$_2$As$_2$ in the TrI state, altering the magnetic order to FM results in a WSM, which induces a insulator-metal transition. Experimentally, such a change can occur in two ways, either by applying a magnetic field to fully polarize the moments [Fig.~\ref{Fig_MR}], or by increasing pressure above $p_{\rm c2}$ [Figs.~\ref{Fig_rho}(c) and (d)], around which an AFM-FM transition occurs (see Fig.~S2 and Section~S2 in the Supplementary Materials) \cite{gati2021pressureinduced}.

Based on the DFT calculations shown in Fig.~\ref{Fig_DFT}, the evolution of the topological classification for EuCd$_2$As$_2$ under pressure is schematically shown in Fig.~\ref{Fig_phase_diagram}(a). Under ambient pressure and inside the AFM state, EuCd$_2$As$_2$ is a small-gap MTI, with topologically-protected surface states contributing to metallic conduction. With increasing pressure, the bulk band inversion necessary for a topological insulator is suppressed and AFM EuCd$_2$As$_2$ becomes a TrI, with diminished charge carriers when cooled to sufficiently low temperatures. When EuCd$_2$As$_2$ goes through an AFM-FM transition, the TrI state evolves to become a WSM, with metallic conduction restored by Weyl fermions and surface Fermi arcs. 

\section*{Discussion}

\begin{figure}
	\includegraphics[width=0.9\columnwidth]{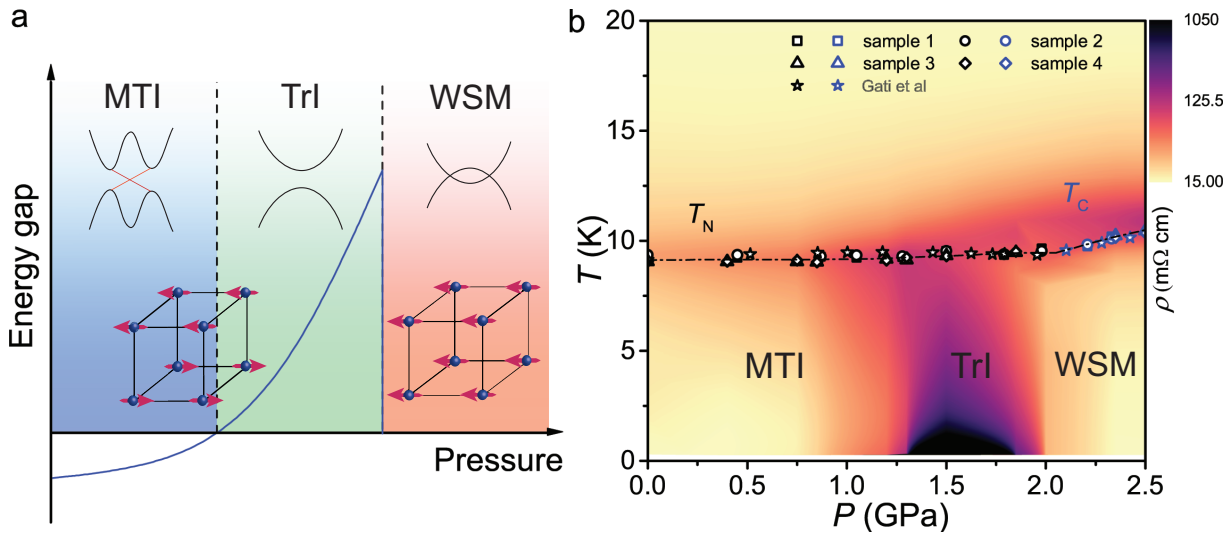} \protect\caption{{\bf Phase diagram of EuCd$_2$As$_2$ under pressure.} (a) Theoretically anticipated evolution of the ground state in EuCd$_2$As$_2$ with increasing pressure, evolving from a magnetic topological insulator (MTI) to a trivial insulator (TrI), and then to a Weyl semimetal (WSM). Magnetic ground states and electronic structures associated with the MTI, TrI and WSM phases are shown schematically. The solid curve represents the evolution of the electronic gap, which is negative in the presence of a band inversion, and positive in its absence. (b) Color-coded experimental $\rho(T)$ as a function of pressure and temperature for sample~\#4 (the results for other samples are provided in the Supplementary Materials, see Fig.~S7), note that $\rho(T)$ is shown on a log scale. 
	$T_{\rm N}$ (black symbols) and $T_{\rm C}$ (blue symbols) mark local maxima in $d\rho/dT$, respectively corresponding to transition temperatures into the AFM and FM phases (see Fig.~S8 in the Supplementary Materials). For comparison, the magnetic transition temperatures from Ref.~\cite{gati2021pressureinduced} are also included in the plot. 
	}
	\label{Fig_phase_diagram}
\end{figure}

Our resistivity measurements under pressure provide compelling evidence for an unprecedented insulating dome inside the AFM state of EuCd$_2$As$_2$ for pressures between $p_{\rm c1}\sim1.0$~GPa and $p_{\rm c2}\sim2.0$~GPa, with an AFM-FM transition in the magnetic ground state also occurring around $p_{\rm c2}$ [Fig.~\ref{Fig_phase_diagram}(b)]. Such an evolution under pressure can be qualitatively captured by the DFT calculations, which suggest that EuCd$_2$As$_2$ goes through consecutive topological phase transitions. The first transition is from a MTI to a TrI at $p_{\rm c1}$, occurring within the A-type AFM phase. The second transition that occurs at $p_{\rm c2}$ is from a TrI to a WSM, driven by the AFM-FM transition in the magnetic ground state. Given the proximity between the AFM and FM ground states \cite{Krishna2018}, the AFM-FM transition can also be driven by an applied field, which accounts for the colossal MR achievable with a small magnetic field. In such a scenario, similar to increasing pressure above $p_{\rm 2}$, the MR is caused by a change from a TrI to a WSM, which occurs both when FM moments are oriented along the $c$-axis or in the $ab$-plane (see Section~S3, Table.~S1 and Figs.~S3-6 in the Supplementary Materials).

The colossal negative MR observed in EuCd$_2$As$_2$ for $p_{\rm c1}<p<p_{\rm c2}$ is unusual, as its magnitude is large, persists to the lowest temperatures, and likely arises from a field-induced topological phase transition from a TrI to a topologically nontrivial WSM. Given that the MR arises from an insulator-metal transition, record-breaking values of MR may be achieved upon cooling to ever lower temperatures. The small fields required to achieve the large values of MR further makes EuCd$_2$As$_2$ appealing for technological applications at very low temperatures. The observation of colossal MR for small fields both along the $c$-axis and in the $ab$-plane [Table~\ref{table_MR}] means that the colossal MR would be robust regardless of the sample orientation, and would persist even in polycrystalline samples. While a modest pressure is required to access the TrI phase of EuCd$_2$As$_2$ with the colossal MR, it may be possible to replace the hydrostatic pressure with chemical pressure via P-substitution, and realize a similar colossal MR under ambient pressure.     

Whereas the transition from a MTI to TrI under pressure within the AFM state of EuCd$_2$As$_2$ arises from enhanced electronic hopping that suppresses the band inversion, both the pressure- and field-induced transitions from a TrI to a WSM requires a transformation of the magnetic ground state. To facilitate such a change, the distinct magnetic ground states should be nearly degenerate, which in the case of EuCd$_2$As$_2$ results from weak magnetic exchange couplings along the $c$-axis, which can be easily overcome by an applied magnetic field. Furthermore, EuCd$_2$As$_2$ exhibits weak magnetic anisotropy due to the vanishing orbital moments of the Eu$^{2+}$ ions, which allows the magnetic moments to become fully polarized regardless of the field orientation, leading to a robust colossal MR that is nearly independent of the relative orientation between the sample and the applied field.
In combination with an electronic topology that depends on the magnetic structure, the weak interlayer magnetic exchange couplings and magnetic anisotropy in EuCd$_2$As$_2$ enables the switching between magnetic ground states, and hence the electronic topology, leading to dramatic changes in the macroscopic electrical transport. These findings demonstrate a clear example of topological phase transitions driven by the manipulation of the magnetic ground state, and underscore weak magnetic exchange couplings and magnetic anisotropy as key ingredients in the search for tunable magnetic topological materials.


\section*{Materials and Methods}
\subsection*{Experimental details}
Single crystals of EuCd$_2$As$_2$ were grown using the Sn-flux method, with details in Ref.~\cite{Sun2021}. Electrical resistivity measurements under pressure were carried out in a piston-cylinder pressure cell using the standard four probe method, with the current in the $ab$-plane. To ensure hydrostaticity, Daphne 7373 was used as the pressure-transmitting medium. Values of the applied pressure were determined from the shift of $T_{\rm c}$ for a high-quality Pb single crystal. All the resistivity measurements were performed down to 0.3~K in a $^3$He refrigerator with a 15~T magnet. Specific heat under ambient pressure was measured using a Quantum Design Physical Property Measurement System (PPMS), using a standard pulse relaxation method. Magnetization measurements were carried out using the vibrating sample magnetometer (VSM) option in a Quantum Design PPMS.

\subsection*{First-principles electronic structure calculations}
The electronic structures of EuCd$_2$As$_2$ under pressure were calculated from first principles using density functional theory (DFT). The DFT calculations were performed using the plane-wave projected augmented wave method as implemented in the \texttt{VASP} code \cite{PhysRevB.47.558,PhysRevB.59.1758}. The plane-wave basis energy cut-off was set to 480~eV, and a $12\times12\times4$ $\Gamma$-centered $k$-mesh was used to perform integration over the Brillouin zone. An additional on-site Coloumb interaction of $U = 6$~eV was included for the Eu-$4f$ orbitals in the LDA+$U$ calculations. Both the lattice constants and the atomic internal coordinates were optimized for all magnetic configurations, so that forces on each atom were smaller than 0.01~eV/{\AA}. The band structures from \texttt{VASP} were fit to tight-binding Hamiltonians using the maximally projected Wannier function method. The resulting tight-binding Wannier-orbital-based Hamiltonians were 
used to calculate the corresponding surface states \cite{1985} and analyze their electronic topology with the \texttt{WannierTools} package \cite{WU2018405}.
The band structures from \texttt{VASP} were fit to tight-binding Hamiltonians using the maximally projected Wannier function method, and their topology were analyzed with the \texttt{WannierTools} package \cite{WU2018405}.  

\section*{Acknowledgments}

We acknowledge helpful discussions with Zhentao Wang, Wei Zhu, and Yang Liu. This work was supported by the National Key R\&D Program of China (No. 2017YFA0303100), the National Natural Science Foundation of China (No. 11974306, No. 12034017 and No. 11874137), and the Key R\&D Program of Zhejiang Province, China (2021C01002).  
\section*{Author contributions}
F.D., L.Y., Z.N., S.L., Y.C. and D.S. carried out the experimental measurements. Y.L. and Y.Shi prepared the samples. F.D., N.W. and C.C performed the first-principles calculations. L.Y. initiated this study and H.Y. supervised the project. F.D., C.C., Y.Song and H.Y. analyzed the results. Y.Song, H.Y., C.C., M.S., and F.S. wrote the manuscript, with inputs from all authors.

\subsection*{Competing interests}
The authors declare no competing interests.

\subsection*{Data and materials availability} 
All data needed to evaluate the conclusions in the paper are present in the paper and/or the Supplementary Materials.

\bibliography{bibfile}
\end{document}